\documentclass[prl,twocolumn,showpacs,superscriptaddress,preprintnumbers,floatfix]{revtex4}

\usepackage{ifpdf}
\usepackage{hyperref}
\usepackage{dcolumn}
\usepackage{url}
\usepackage{amsmath}
\usepackage{amscd}
\usepackage{amsfonts}
\usepackage{amssymb}
\usepackage{bm}   
\usepackage{bbm}   
\usepackage{verbatim}
\usepackage{stmaryrd}
\usepackage{amsthm}

\ifpdf
  \usepackage[pdftex]{graphicx}         
\else
  \usepackage[dvips]{graphicx}
\fi

\theoremstyle{plain}   \newtheorem{Lem}{Lemma}
\theoremstyle{plain} 	\newtheorem{Cor}{Corollary}
\theoremstyle{plain} 	\newtheorem{The}{Theorem}
\theoremstyle{plain} 	\newtheorem{Prop}{Proposition}
\theoremstyle{plain} 	
\theoremstyle{plain}	
\theoremstyle{plain}	\newtheorem*{Def}{Definition} 
\theoremstyle{plain}	

\newcommand{\eqnref}[1]{Eq.~(\ref{#1})}
\newcommand{\Eqnref}[1]{Equation~(\ref{#1})}
\newcommand{\figref}[1]{Fig.~\ref{#1}}

\newcommand{\theref}[1]{Thm.~\ref{#1}}

\newcommand{\propref}[1]{Prop.~\ref{#1}}

\newcommand{\corref}[1]{Cor.~\ref{#1}}

\newcommand{\lemref}[1]{Lem.~\ref{#1}}

\newcommand{\refcite}[1]{Ref.~\cite{#1}}
\newcommand{\Refcite}[1]{Reference~\cite{#1}}

\newcommand{\order}[1]{order\protect\nobreakdash-$#1$}

\newcommand{\cryptic}[1]{$#1$\protect\nobreakdash-cryptic}
\newcommand{\Cryptic}[1]{$#1$\protect\nobreakdash-Cryptic}
\newcommand{\crypticity}[1]{$#1$\protect\nobreakdash-crypticity}

\newcommand{\eM}     {$\epsilon$\protect\nobreakdash-machine}
\newcommand{\eMs}    {$\epsilon$\protect\nobreakdash-machines}

\newcommand{\EMs}    {$\epsilon$\protect\nobreakdash-Machines}

\newcommand{\Process}{\mathcal{P}}

\newcommand{\MeasAlphabet}	{\mathcal{A}}
\newcommand{\MeasSymbol}   { {X} }
\newcommand{\meassymbol}   { {x} }

\newcommand{\Past}	{ \overleftarrow {\MeasSymbol} }
\newcommand{\past}	{ {\overleftarrow {\meassymbol}} }

\newcommand{\Future}	{ \overrightarrow{\MeasSymbol} }
\newcommand{\future}	{ \overrightarrow{\meassymbol} }

\newcommand{\CausalState}	{ \mathcal{S} }
\newcommand{\CausalStatePrime}	{ {\CausalState}^{\prime}}
\newcommand{\causalstate}	{ \sigma }
\newcommand{\causalstateprime}	{ \sigma^{\prime} }
\newcommand{\CausalStateSet}	{ \boldsymbol{\CausalState} }

\newcommand{\Cmu}		{C_\mu}
\newcommand{\hmu}		{h_\mu}
\newcommand{\EE}		{{\bf E}}

\newcommand{\PC}		{\chi}
\newcommand{\FuturePC}		{\PC^+}
\newcommand{\PastPC}		{\PC^-}

\newcommand{\forward}{+}
\newcommand{\reverse}{-}

\newcommand{\FutureCausalState}	{ {\CausalState}^{\forward} }

\newcommand{\PastCausalState}	{ {\CausalState}^{\reverse} }

\newcommand{\FutureCmu}	{ C_\mu^{\forward} }
\newcommand{\PastCmu}	{ C_\mu^{\reverse} }

\newcommand{\one}{\mathbf{1}}

\begin{document}

\title{Information Accessibility and Cryptic Processes}

\author{John R. Mahoney}
\email{jrmahoney@@ucdavis.edu}
\affiliation{Complexity Sciences Center and Physics Department,
University of California at Davis, One Shields Avenue, Davis, CA 95616}

\author{Christopher J. Ellison}
\email{cellison@@cse.ucdavis.edu}
\affiliation{Complexity Sciences Center and Physics Department,
University of California at Davis, One Shields Avenue, Davis, CA 95616}

\author{James P. Crutchfield}
\email{chaos@@cse.ucdavis.edu}
\affiliation{Complexity Sciences Center and Physics Department,
University of California at Davis, One Shields Avenue, Davis, CA 95616}
\affiliation{Santa Fe Institute, 1399 Hyde Park Road, Santa Fe, NM 87501}

\date{\today}

\bibliographystyle{unsrt}

\begin{abstract}
We give a systematic expansion of the crypticity---a recently introduced
measure of the inaccessibility of a stationary process's internal state
information. This leads to a hierarchy of \cryptic{k} processes and allows us
to identify finite-state processes that have infinite crypticity---the internal
state information is present across arbitrarily long, observed sequences. The
crypticity expansion is exact in both the finite- and infinite-order cases. It
turns out that \crypticity{k} is complementary to the Markovian finite-order
property that describes state information in processes. One application of
these results is an efficient expansion of the excess entropy---the mutual
information between a process's infinite past and infinite future---that is
finite and exact for finite-order cryptic processes.
\end{abstract}

\preprint{Santa Fe Institute Working Paper 09-05-XXX}
\preprint{arxiv.org:0905.XXXX [physics.cond-mat]}

\maketitle


\section{Introduction}

The data of phenomena come to us through observation. A large fraction of the
theoretical activity of model building, though, focuses on internal mechanism.
How are observation and modeling related? A first step is to frame the problem
in terms of hidden processes---internal mechanisms probed via instruments that,
in particular, need not accurately report a process's internal state. A
practical second step is to measure the difference between internal structure
and the information in observations.

We recently established that the amount of observed information a process
communicates from the past to the future---the \emph{excess entropy}---is the
mutual information between its forward- and reverse-time minimal causal
representations \cite{Crut08a,Crut08b}. This closed-form expression gives a
concrete connection between the observed information and a process's internal
structure.

Excess entropy, and related mutual information quantities, are widely used
diagnostics for complex systems. They have been applied to detect the presence
of organization in dynamical systems~\cite{Fras90b,Casd91a,Spro03a,Kant06a},
in spin systems~\cite{Arno96,Crut97a,Feld98b}, in neurobiological systems~%
\cite{Tono94a,Bial00a}, and even in language~\cite{Ebel94c,Debo08a}, to mention
only a very few uses. Thus, understanding how much internal state structure is
reflected in the excess entropy is critical to whether or not these and other
studies of complex systems can draw structural inferences about the internal
mechanisms that produce observed behavior.

Unfortunately, there is a fundamental problem. The excess entropy is \emph{not}
the internal state information the process stores---rather, the latter is the
process's \emph{statistical complexity} \cite{Crut08a,Crut08b}. On the positive
side, there is a diagnostic. The difference between, if you will, experiment and
theory (between observed information and internal structure) is controlled by
the difference between a process's excess entropy and its statistical
complexity. This difference is called the \emph{crypticity}---how much internal
state information is inaccessible \cite{Crut08a,Crut08b}. Here we introduce a
classification of processes using a systematic expansion of crypticity.

The starting point is \emph{computational mechanics}'s minimal causal
representation of a stochastic process $\Process$---the \emph{\eM}
\cite{Crut88a,Crut98d}.
There, a process is viewed as a channel that communicates information from the
past, $\Past = \ldots \MeasSymbol_{-3} \MeasSymbol_{-2} \MeasSymbol_{-1}$, to
the future, $\Future = \MeasSymbol_0 \MeasSymbol_1 \MeasSymbol_2 \ldots$.
($\MeasSymbol_t$ takes values in a finite measurement alphabet $\MeasAlphabet$.)
The excess entropy is the shared (or mutual) information between
the past and the future: $\EE = I[\Past;\Future]$. The amount of historical
information that a process stores in the present is different. It is given by
the Shannon information $\Cmu = H[\CausalState]$ of the distribution over the
\eM's \emph{causal states} $\CausalStateSet$. $\Cmu$ is called the
\emph{statistical complexity} and the causal states are sets of pasts $\past$
that are equivalent for prediction \cite{Crut88a}:
\begin{equation}
\epsilon(\past) =
  \{ \past^\prime: \Pr(\Future|\past) = \Pr(\Future|\past^\prime) \} ~.
\label{Eq:PredictiveEquivalence}
\end{equation}
Causal states have a Markovian property that they render the past and future
statistically independent; they \emph{shield} the future from the
past~\cite{Crut98d}:
\begin{equation}
\Pr(\Past,\Future|\CausalState)
  = \Pr(\Past|\CausalState) \Pr(\Future|\CausalState) ~.
\label{shield}
\end{equation}
\EMs\ are also \emph{unifilar}~\cite{Crut88a,Shal98a}: From the start state,
each observed sequence $\ldots x_{-3} x_{-2} x_{-1} \ldots$ corresponds to one
and only one sequence of causal states. The signature of unifilarity is that
on knowing the current state and measurement, the uncertainty in the next
state vanishes: $H[\CausalState_{t+1}|\CausalState_t,\MeasSymbol_t] = 0$.

Although they are not the same, the basic relationship between these quantities
is clear: $\EE$ is the process's effective channel utilization and $\Cmu$ is
the sophistication of that channel. Their difference, one of our main concerns
in the following, indicates how a process stores, manipulates, and hides
internal state information.

Until recently, $\EE$ could not be as directly calculated from the \eM\ as the
process's entropy rate $\hmu$ and its statistical complexity. \refcite{Crut08a}
and \refcite{Crut08b} solved this problem, giving a closed-form expression for
the excess entropy:
\begin{equation}
\EE = I[\FutureCausalState;\PastCausalState] ~,
\label{eq:EForRevState}
\end{equation}
where $\FutureCausalState$ are the causal states of the process scanned in
the ``forward'' direction and $\PastCausalState$ are the causal states of
the process scanned in the ``reverse'' time direction.

This result comes in a historical context. Some time ago, an explicit
expression for the excess entropy had been developed from the Hamiltonian
for one-dimensional spin chains with range-$R$ interactions~\cite{Crut97a}:
\begin{equation}
\EE = \Cmu - R \, \hmu ~.
\label{eq:SpinEEandCmu}
\end{equation}
A similar, but slightly less compact form is known for \order{R} Markov
processes:
\begin{equation}
\EE = H[\MeasSymbol_0^R] - R \, \hmu ~,
\label{eq:MarkovEE}
\end{equation}
where $\MeasSymbol_0^R = X_0 , \ldots, X_{R-1}$.
It has also been known for some time that the statistical complexity
is an upper bound on the excess entropy~\cite{Shal98a}:
\begin{align*}
\EE \leq \Cmu ~,
\end{align*}
which follows from the equality derived there:
\begin{align*}
\EE = \Cmu - H[\FutureCausalState|\Future] ~.
\label{eq:OldECmu}
\end{align*}

Using forward and reverse \eMs, \refcite{Crut08a} extended this, deriving the
closed-form expression for $\EE$ in \eqnref{eq:EForRevState} and two new
bounds on $\EE$: $\EE \leq \PastCmu$ and $\EE \leq \FutureCmu$. It also showed
that:
\begin{align}
H[\FutureCausalState|\Future] = H[\FutureCausalState|\PastCausalState]
\end{align}
and identified this quantity as controlling how a process hides its internal
state information. For this reason, it is called the process's
\emph{crypticity}:
\begin{equation}
\FuturePC = H[\FutureCausalState|\Future] ~.
\label{eq:crypticitydef}
\end{equation}
In the context of forward and reverse \eM s, one must distinguish
two crypticities; depending on the scan direction one has:
\begin{align*}
\FuturePC & = H[\FutureCausalState|\PastCausalState] \text{ or}\\
\PastPC   & = H[\FutureCausalState|\PastCausalState] ~.
\end{align*}
In the following we will not concern ourselves with reverse representations
and so can simplify the notation, using $\Cmu$ for $\FutureCmu$ and $\PC$ for
$\FuturePC$.

Here we show that, for a restricted class of processes, the crypticity in
\eqnref{eq:OldECmu} can be systematically expanded to give an alternative
closed-form to the excess entropy in \eqnref{eq:EForRevState}. One ancillary
benefit is a new and, we argue, natural hierarchy of processes in terms of
information accessibility.

\section{k-Crypticity}

The process classifications based on spin-block length and \order{R} Markov
are useful. They give some insight into the nature of the kinds of process we
can encounter and, concretely, they allow for closed-form expressions for the
excess entropy (and other system properties). In a similar vein, we wish to
carve the space of processes with a new blade. We define the class of
\emph{\cryptic{k}} processes and develop their properties and closed-form
expressions for their excess entropies.

For convenience, we need to introduce several shorthands. First, to denote a
symbol sequence that begins at time $t$ and is $L$ symbols long, we write
$\MeasSymbol_t^L$. Note that $\MeasSymbol_t^L$ includes $\MeasSymbol_{t+L-1}$,
but not $\MeasSymbol_{t+L}$. Second, to denote a symbol sequence that begins at
time $t$ and continues on to infinity, we write $\Future_t$.

\begin{Def}
The \emph{\crypticity{k} criterion} is satisfied when
\begin{align}
H[\CausalState_k | \Future_0] = 0 ~.
\label{Def:OrderKCrypticityCriterion}
\end{align}
\end{Def}

\begin{Def}
A \emph{\cryptic{k}} process is one for which the process's \eM\ satisfies
the \crypticity{k} criterion.
\end{Def}

\begin{Def}
An \emph{\cryptic{\infty}} process is one for which the process's \eM\ does
not satisfy the \crypticity{k} criterion for any finite $k$.
\end{Def}

\begin{Lem}
\label{lem:hSkgF}
$H[\CausalState_k | \Future_0]$ is a nonincreasing function of $k$.
\end{Lem}

\begin{proof}
This follows directly from stationarity and the fact that
conditioning on more random variables cannot increase entropy:
\begin{align*}
H[\CausalState_{k+1} | \Future_0]
    = [\CausalState_k | \Future_{-1}]
    \leq H[\CausalState_k | \Future_0] ~.
\end{align*}
\end{proof}

\begin{Lem}
If $\Process$ is \cryptic{k},
then $\Process$ is also \cryptic{j} for all $j > k$.
\label{lem:OKCOJC}
\end{Lem}

\begin{proof}
Being \cryptic{k} implies $H[\CausalState_k | \Future_0] = 0$.
Applying \lemref{lem:hSkgF},
$H[\CausalState_j | \Future_0] \le H[\CausalState_k | \Future_0] = 0$.
By positivity of entropy, we conclude that $\Process$ is also \cryptic{j}.
\end{proof} 

This provides us with a new way of partitioning the space of processes. We
create a parametrized class of sets $\{\chi_k: k = 0,1,2, \ldots\}$, where
$\chi_k = \{\Process: \text{ \cryptic{k} and not \cryptic{(k-1)}}\}$.

The following result provides a connection to a very familiar class of
processes.

\begin{Prop}
If a process $\Process$ is \order{k} Markov, then it is \cryptic{k}.
\label{prop:MarkovOKC}
\end{Prop}

\begin{proof}
If $\Process$ is \order{k} Markov, then $H[\CausalState_k | \MeasSymbol_0^k]=0$.
Conditioning on more variables does not increase uncertainty, so:
\begin{equation*}
H[\CausalState_k | \MeasSymbol_0^k, \Future_k] = 0 ~.
\end{equation*}
But the lefthand side is $H[\CausalState_k | \Future_0]$. Therefore,
$\Process$ is \cryptic{k}.
\end{proof}

Note that the converse of Prop. \ref{prop:MarkovOKC} is not true. For example,
the Even Process (EP), the Random Noisy Copy Process (RnC), and the Random
Insertion Process (RIP) (see \refcite{Crut08a} and \refcite{Crut08b}), are all
\cryptic{1}, but are not \order{R} Markov for any finite $R$.

Note also that Prop.~\ref{prop:MarkovOKC} does not preclude an \order{k}
Markov process from being \cryptic{j}, where $j < k$.
Later we will show an example demonstrating this.

Given a process, in general one will not know its crypticity order. One way to
investigate this is to study the sequence of estimates of $\PC$ at different
orders. To this end, we define the \cryptic{k} approximation.

\begin{Def}
The \emph{\cryptic{k} approximation} is defined as
\begin{align*}
\PC(k) = H[\CausalState_0 | \MeasSymbol_0^k, \CausalState_k] ~.
\end{align*}
\end{Def}

\subsection{The \Cryptic{k} Expansion}

We will now develop a systematic expansion of $\PC$ to order $k$ in which
$\PC(k)$ appears directly and the \crypticity{k} criterion plays the
role of an error term.

\begin{The}
The process crypticity is given by
\begin{align}
\PC = \PC(k) + H[\CausalState_k | \Future_0] ~.
\label{eq:CrypticityExpansion}
\end{align}
\label{thm:OKCC}
\end{The}

\begin{proof}
We calculate directly, starting from the definition, adding and subtracting
the \crypticity{k} criterion term from $\PC$'s definition,
\eqnref{eq:crypticitydef}:
\begin{align*}
\PC = H[\CausalState_0 | \Future_0] - H[\CausalState_k | \Future_0]
		+ H[\CausalState_k | \Future_0] ~.
\end{align*}
We claim that the first two terms are $\PC(k)$. Expanding the conditionals in
the purported $\PC(k)$ terms and then canceling, we get joint distributions:
\begin{align*}
H[\CausalState_0 | \Future_0] - H[\CausalState_k | \Future_0]
= H[\CausalState_0,  \Future_0] - H[\CausalState_k, \Future_0] ~.
\end{align*}
Now, splitting the future into two pieces and using this to write
conditionals, the righthand side becomes:
\begin{align*}
H[\Future_k | \CausalState_0, \MeasSymbol_0^k]
	+ H[\CausalState_0, \MeasSymbol_0^k]
	- H[\Future_k | \CausalState_k, \MeasSymbol_0^k]
	- H[\CausalState_k, \MeasSymbol_0^k] ~.
\end{align*}
Appealing to the \eM's unifilarity, we then have:
\begin{align*}
H[\Future_k | \CausalState_k] + H[\CausalState_0, \MeasSymbol_0^k]
	- H[\Future_k | \CausalState_k, \MeasSymbol_0^k]
	- H[\CausalState_k, \MeasSymbol_0^k] ~.
\end{align*}
Now, applying causal shielding gives:
\begin{align*}
H[\Future_k | \CausalState_k] + H[\CausalState_0, \MeasSymbol_0^k]
	- H[\Future_k | \CausalState_k]
	- H[\CausalState_k, \MeasSymbol_0^k] ~.
\end{align*}
Canceling terms, this simplifies to:
\begin{align*}
H[\CausalState_0, \MeasSymbol_0^k] - H[\CausalState_k, \MeasSymbol_0^k] ~.
\end{align*}
We now re-expand, using unifilarity to give:
\begin{align*}
H[\CausalState_0, \MeasSymbol_0^k, \CausalState_k]
	- H[\CausalState_k, \MeasSymbol_0^k] ~.
\end{align*}
Finally, we combine these, using the definition of conditional entropy, to
simplify again:
\begin{align*}
H[\CausalState_0 | \MeasSymbol_0^k, \CausalState_k] ~.
\end{align*}
Note that this is our definition of $\PC(k)$.

This establishes our original claim:
\begin{align*}
\PC = \PC(k) + H[\CausalState_k | \Future_0] ~,
\end{align*}
with the \crypticity{k} criterion playing the role of an approximation
error.

\end{proof}

\begin{Cor}
A process $\Process$ is \cryptic{k} if and only if
\begin{align*}
\PC = \PC(k) ~.
\end{align*}
\label{cor:OKCC}
\end{Cor}

\begin{proof}
Given the \order{k} expansion of $\PC$ just developed, we now assume the
\crypticity{k} criterion is satisfied; viz.,
$H[\CausalState_k | \Future_0] = 0$. Thus, we have from
\eqnref{eq:CrypticityExpansion}:
\begin{align*}
\PC = \PC(k) ~.
\end{align*}
Likewise, assuming $\PC = \PC(k)$ requires, by \eqnref{eq:CrypticityExpansion} 
that $H[\CausalState_k | \Future_0] = 0$ and thus the process is \cryptic{k}.
\end{proof}

\begin{Cor}
For any process, $\PC(0) = 0$.
\label{cor:ZeroCrypticity}
\end{Cor}

\begin{proof}
\begin{align*}
\PC(0) &= H[\CausalState_0 | \MeasSymbol_0^0, \CausalState_0]\\
  &= H[\CausalState_0 | \CausalState_0] = 0~.
\end{align*}
\end{proof}

\subsection{Convergence}

\begin{Prop}
The approximation $\PC(k)$ is a nondecreasing function of $k$. 
\label{prop:NonDecreasingCrypticity}
\end{Prop}

\begin{proof}
\lemref{lem:hSkgF} showed that $H[\CausalState_k | \Future_0]$ is a
nonincreasing function of $k$.  By \theref{thm:OKCC}, $\PC(k)$ must be a
nondecreasing function of $k$.
\end{proof} 

\begin{Cor}
Once $\PC(k)$ reaches the value $\PC$, $\PC(j) = \PC$ for all $j > k$.
\label{cor:ReachChiStayChi}
\end{Cor}

\begin{proof}
If there exists such a $k$, then by \theref{thm:OKCC} the process is
\cryptic{k}. By \lemref{lem:OKCOJC}, the process is \cryptic{j} for all
$j>k$. Again, by \theref{thm:OKCC}, $\PC(j) = \PC$.
\end{proof}

\begin{Cor}
If there is a $k \geq 1$ for which $\PC(k)=0$, then $\PC(1)=0$.
\label{cor:Chik0Chi10}
\end{Cor}

\begin{proof}
By positivity of the conditional entropy
$H[\CausalState_0 | \MeasSymbol_0, \CausalState_1]$, $\PC(1) \ge 0$. By the
nondecreasing property of $\PC(k)$ from \propref{prop:NonDecreasingCrypticity},
$\PC(1) \le \PC(k) = 0$. Therefore, $\PC(1) = 0$.
\end{proof}

\begin{Cor}
If $\PC(1)=0$, then $\PC(k) = 0$ for all $k$.
\label{cor:Chi10Chik0}
\end{Cor}

\begin{proof}
Applying stationarity, $\PC(1) = H[\CausalState_0|\MeasSymbol_0,\CausalState_1]
  = H[\CausalState_k | \MeasSymbol_k, \CausalState_{k+1}]$. We are given
$\PC(1) = 0$ and so $H[\CausalState_k | \MeasSymbol_k, \CausalState_{k+1}] = 0$.
We use this below. Expanding $\PC(k+1)$,
\begin{align*}
\PC(k+1) &= H[\CausalState_0 | \MeasSymbol_0^{k+1}, \CausalState_{k+1}]\\
&=  H[\CausalState_0 | \MeasSymbol_0^{k}, \MeasSymbol_k, \CausalState_{k+1}]\\
&=  H[\CausalState_0 | \MeasSymbol_0^{k}, \CausalState_k, \MeasSymbol_k, \CausalState_{k+1}]\\
&\le  H[\CausalState_0 | \MeasSymbol_0^{k}, \CausalState_k]\\
&= \PC(k) ~.
\end{align*}
The third line follows from $\PC(1)=0$. 
By \propref{prop:NonDecreasingCrypticity}, $\PC(k+1) \ge \PC(k)$. 
Therefore, $\PC(k+1) = \PC(k)$. Finally, using $\PC(1)=0$, we have by 
induction that $\PC(k)=0$ for all $k$.
\end{proof}


\begin{Cor}
If there is a $k \geq 1$ for which $\PC(k)=0$, then $\PC(j)=0$ for all
$j \geq 1$.
\label{cor:TrivialChi}
\end{Cor}

\begin{proof}
This follows by composing \corref{cor:Chik0Chi10} with \corref{cor:Chi10Chik0}.
\end{proof}


Together, the proposition and its corollaries show that $\PC(k)$ is a
nondecreasing function of $k$ which, if it reaches $\PC$ at a finite $k$,
remains at that value for all larger $k$.

\begin{Prop}
The cryptic approximation $\PC(k)$ converges to $\PC$ as $k \to \infty$.
\label{prop:ChiConverges}
\end{Prop}

\begin{proof}
Note that $\PC = \lim_{k \to \infty}{H[\CausalState_0 | \MeasSymbol_0^k]}$ and
recall that $\PC(k) = H[\CausalState_0 | \MeasSymbol_0^k, \CausalState_k]$.
We show that the difference approaches zero:
\begin{align*}
H[\CausalState_0 | \MeasSymbol_0^k]
  & - H[\CausalState_0 | \MeasSymbol_0^k, \CausalState_k]\\
  & = H[\CausalState_0, \MeasSymbol_0^k] - H[\MeasSymbol_0^k] \\
  & \;\;\;\; - H[\CausalState_0, \MeasSymbol_0^k, \CausalState_k]
	+ H[\MeasSymbol_0^k, \CausalState_k]\\
  & = H[\CausalState_0, \MeasSymbol_0^k] - H[\MeasSymbol_0^k] \\
  & \;\;\;\; - H[\CausalState_0, \MeasSymbol_0^k]
	+ H[\MeasSymbol_0^k, \CausalState_k]\\
  & = H[\MeasSymbol_0^k, \CausalState_k] - H[\MeasSymbol_0^k]\\
  & = H[\CausalState_k | \MeasSymbol_0^k] ~.
\end{align*}
Moreover, $\lim_{k \to \infty}{H[\CausalState_k | \MeasSymbol_0^k]} = 0$ by the
$\epsilon$ map from pasts to causal states of \eqnref{Eq:PredictiveEquivalence}.
Therefore, as $k \to \infty$, $\PC(k) \to \PC$.
\end{proof}

\subsection{Excess Entropy for \Cryptic{k} Processes}

Given a \cryptic{k} process, we can calculate its excess entropy in a form
that involves a sum of $\propto |\mathcal{A}^k|$ terms, where each term
involves products of $k$ matrices. Specifically, we have the following.

\begin{Cor}
A process $\Process$ is \cryptic{k} if and only if $\EE = \Cmu - \PC(k)$.
\label{cor:OKCE}
\end{Cor}

\begin{proof}
From \refcite{Crut08a}, we have $\EE = \Cmu - \PC$, and by \corref{cor:OKCC},
$\PC = \PC(k)$. Together, these complete the proof.
\end{proof}

The following proposition is a simple and useful consequence of the class
of \cryptic{k} processes.

\begin{Cor}
A process $\Process$ is \cryptic{0} \emph{if} and \emph{only if} $\EE = \Cmu$.
\label{cor:EEEqualCmu}
\end{Cor}

\begin{proof}
If $\Process$ is \cryptic{0}, our general expression then reads
\begin{align*}
\EE &= \Cmu - H[\CausalState_0 | \MeasSymbol_0^0, \CausalState_0]\\
  &= \Cmu ~.
\end{align*}
To establish the opposite direction, $\EE = \Cmu$ and \corref{cor:OKCE} imply
that $\PC(k) = 0$ for all $k$. In particular, $\PC(0)$ and the process is
\cryptic{0}.
\end{proof}

\subsection{Crypticity versus Markovity}

\Eqnref{eq:SpinEEandCmu} and \Eqnref{eq:MarkovEE} give expressions for $\EE$
in the cases when the process is \order{R} Markov and when it is an \order{R}
spin chain. These results hinge on whether or not $H[\MeasSymbol_0^R] = \Cmu$.

Reference~\cite{Crut97a} stated a condition under which equality holds in terms
of transfer matrices. Here we state a simpler condition by equating two chain
rule expansions of $H[\MeasSymbol_0^R, \CausalState_R]$:
\begin{align*}
H[\MeasSymbol_0^R | \CausalState_R] + H[\CausalState_R]
= H[\CausalState_R | \MeasSymbol_0^R] + H[\MeasSymbol_0^R] ~.
\end{align*}
$H[\CausalState_R | \MeasSymbol_0^R] = 0$ by virtue of the fact that each such
(history) word maps to exactly one causal state by
\eqnref{Eq:PredictiveEquivalence}. Thus, we conclude that for \order{R}
Markov processes:
\begin{align*}
H[\MeasSymbol_0^R] = H[\CausalState_R] \quad \iff \quad 
  H[\MeasSymbol_0^R | \CausalState_R] = 0 ~.
\end{align*}
So, an \order{R} Markov process is also a spin chain \emph{if} 
and \emph{only if} $H[\MeasSymbol_0^R | \CausalState_R] = 0$. 
This means that there is a $1-1$ correspondence between the 
$R$-blocks and causal states, confirming the interpretation specified in 
Ref.~\cite{Crut97a}.

We can also extend the condition for $H[\MeasSymbol_0^R] = \Cmu$ to the
results presented here in the following way.
\begin{Prop}
\begin{align}
H[\MeasSymbol_0^R | \CausalState_R] = 0 \quad \iff \quad
   \PC(R) = R \, \hmu ~,
\end{align}
where $\hmu$ is the process's entropy rate.
\label{prop:SpinChainCondition}
\end{Prop}

\begin{proof}
The proof is a direct calculation:
\begin{align*}
\PC(R) & = H[\CausalState_0 | \MeasSymbol_0^R, \CausalState_R] \\
  & = H[\CausalState_0, \MeasSymbol_0^R] - H[\MeasSymbol_0^R, \CausalState_R] \\
  & = H[\CausalState_0, \MeasSymbol_0^R] - H[\MeasSymbol_0^R | \CausalState_R]
  	- H[\CausalState_R] \\
  & = H[\CausalState_0, \MeasSymbol_0^R] - H[\MeasSymbol_0^R | \CausalState_R]
  	- H[\CausalState_0] \\
  & = H[\MeasSymbol_0^R | \CausalState_0] - H[\MeasSymbol_0^R | \CausalState_R] \\
  & = R \hmu - H[\MeasSymbol_0^R | \CausalState_R] ~.
\end{align*}
\end{proof}

\begin{Prop}
Periodic processes can be arbitrary \order{R} Markov, but are all \cryptic{0}.
\end{Prop}

\begin{proof}
According to \refcite{Crut01a}, we have $\EE = \Cmu$. By \corref{cor:EEEqualCmu}
the process is \cryptic{0}.
\end{proof}

\begin{Prop}
A positive entropy-rate process that is an \order{R} Markov spin chain is not
\cryptic{(R-1)}.
\label{prop:SpinChainNotRMinus1}
\end{Prop}

\begin{proof}
Assume that the \order{R} Markov spin chain \emph{is} \cryptic{(R-1)}.

For $R \ge 1$, If the process is \cryptic{(R-1)}, then by \corref{cor:OKCC}
$\PC(R-1) = \PC$.  Combining this with the above
\propref{prop:SpinChainCondition}, we have
$\PC(R-1) = (R-1) \hmu - H[\MeasSymbol_0^{R-1} | \CausalState_{R-1}]$. If it is
an \order{R} Markov spin chain, then we also have from \eqnref{eq:SpinEEandCmu}
that $\PC = R \hmu$. Combining this with the previous equation, we find that
$H[\MeasSymbol_0^{R-1} | \CausalState_{R-1}] = -\hmu$. By positivity of
conditional entropies, we have reached a contradiction. Therefore an \order{R}
Markov spin chain must not be \cryptic{(R-1)}.

For $R=0$, the proof also holds since negative cryptic orders are not defined.
\end{proof}

\begin{Prop}
A positive entropy-rate process that is an \order{R} Markov spin chain is not
\cryptic{(R-n)} for any $1 \ge n \ge R$.
\end{Prop}

\begin{proof}
For $R \ge 1$,
By \lemref{lem:OKCOJC}, if the process were \cryptic{(R-n)} for some
$1 \ge n \ge R$, then it would be \cryptic{(R-1)}.
By \propref{prop:SpinChainNotRMinus1}, this is not true. Therefore, the
primitive orders of Markovity and crypticity are the same. Similarly, for
$R=0$, the proof also holds since negative cryptic orders are not defined.
\end{proof}

\section{Examples}

It is helpful to see crypticity in action. We now turn to a number of examples
to illustrate how various orders of crypticity manifest themselves in
\eM\ structure and what kinds of processes are cryptic and so hide internal
state information from an observer. For details (transition matrices, notation,
and the like) not included in the following and for complementary discussions
and analyses of them, see Refs. \cite{Crut01a,Crut08a,Crut08b}.

We start at the bottom of the crypticity hierarchy with a \cryptic{0} process
and then show examples of \cryptic{1} and \cryptic{2} processes. Continuing up
the hierarchy, we generalize and give a parametrized family of processes that
are \cryptic{k}. Finally, we
demonstrate an example that is \cryptic{\infty}.

\subsection{Even Process: \Cryptic{0}}

\begin{figure}[th]
\centering
\includegraphics{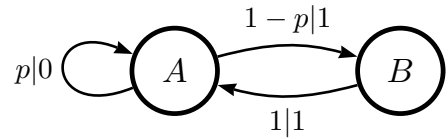}
\caption{A \cryptic{0} process: Even Process. The transitions denote the
	probability $p$ of generating symbol $x$ as $p|x$.}
\label{fig:EvenProcess}
\end{figure}

Figure~\ref{fig:EvenProcess} gives the \eM\ for the Even Process. The Even
Process produces binary sequences in which all blocks of uninterrupted
$1$s are even in length, bounded by zeros. Further, after each even length
is reached, there is a probability $p$ of breaking the block of $1$s by
inserting one or more $0$s.

\Refcite{Crut08b} showed that the Even Process is \cryptic{0} with a
statistical complexity of $\Cmu = H \left( 1/(2-p) \right)$, an entropy rate of
$\hmu = H(p)/(2-p)$, and crypticity of $\PC = 0$. If $p = \frac{1}{2}$, then
$\Cmu = \log_2 (3)-\frac{2}{3}$ bits and $\EE = \log_2 (3)-\frac{2}{3}$ bits.
(As \refcite{Crut08b} notes, these closed-form expressions for $\Cmu$ and
$\EE$ have been known for some time.)

To see why the Even Process is \cryptic{0}, note that if $\MeasSymbol=0$,
then $\CausalState_0=A$; and if $\MeasSymbol=1$, then $\CausalState_0=B$.
Therefore, the \crypticity{0} criterion of
Eq.~(\ref{Def:OrderKCrypticityCriterion}) is satisfied.

It is important to note that this process is \emph{not} \order{R} Markov for any
finite $R$ \cite{Crut01a}. Nonetheless, our new expression for $\EE$ is valid.
This shows the broadening of our ability to calculate $\EE$ even for low
complexity processes that are, in effect, infinite-order Markov.

\subsection{Golden Mean Process: \Cryptic{1}}

\begin{figure}[th]
\centering
\includegraphics{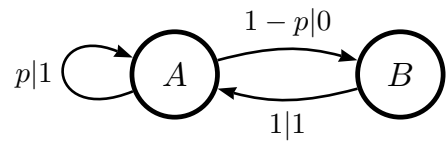}
\caption{A \cryptic{1} process: Golden Mean Process.
  }
\label{fig:GoldenMean}
\end{figure}

Figure~\ref{fig:GoldenMean} shows the \eM\ for the Golden Mean Process
\cite{Crut01a}.
The Golden Mean Process is one in which no two $0$s occur consecutively. After
each $1$, there is a probability $p$ of generating a $0$. As sequence length
grows, the ratio of the number of allowed words of length $L$ to the number
of allowed words at length $L-1$ approaches the golden ratio; hence, its name.
The Golden Mean Process \eM\ looks remarkably similar to that for the Even
Process. The informational analysis, however, shows that they have markedly
different properties.

Reference \cite{Crut08b} showed that the Golden Mean Process has the same
statistical complexity and entropy rate as the Even Process:
$\Cmu = H \left( 1/(2-p) \right)$ and $\hmu = H(p)/(2-p)$. However, the
crypticity is not zero (for $0 < p < 1$). From Cor.~\ref{cor:OKCC} we calculate:
\begin{align*}
\PC & = \PC(1) \\
    & = H[\CausalState_0 | \MeasSymbol_0^1, \CausalState_1] \\
	& = H[\CausalState_0 | \MeasSymbol_0^1]\\
	& = Pr(0) H[\CausalState_0 | \MeasSymbol_0 = 0]
		+ Pr(1) H[\CausalState_0 | \MeasSymbol_0 = 1]\\
    & = H(p)/(2-p) ~.
\end{align*}
If $p = \frac{1}{2}$, $\Cmu = \log_2 (3)-\frac{2}{3}$ bits, an excess
entropy of $\EE = \log_2 (3)-\frac{4}{3}$ bits, and a crypticity
of $\PC = \frac{2}{3}$. Thus, the excess entropy
differs from that of the Even Process. (As with
the Even Process, these closed-form expressions for $\Cmu$ and $\EE$ have been
known for some time.)

The Golden Mean Process is \cryptic{1}. To see why, it is enough to note
that it is \order{1} Markov. By \propref{prop:MarkovOKC}, it is \cryptic{1}.
We know it is not \cryptic{0} since any future beginning with 1 could 
have originated in either state A or B. In addition, the spin-block expression
for excess entropy of \refcite{Crut01a}, \eqnref{eq:SpinEEandCmu} here,
applies for an $R = 1$ Markov chain.

\subsection{Butterfly Process: \Cryptic{2}}

\begin{figure}[th]
\centering
\includegraphics{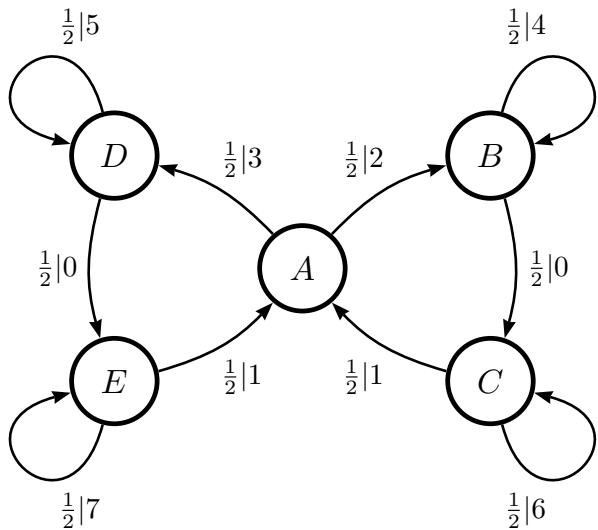}
\caption{A \cryptic{2} process: Butterfly Process over a $6$-symbol alphabet.}
\label{fig:ButterflyProcess}
\end{figure}

The next example, the Butterfly Process of Fig.~\ref{fig:ButterflyProcess},
illustrates in a more explicit way
than possible with the previous processes the role that crypticity plays and
how it can be understood in terms of an \eM's structure. Much of the
explanation does not require calculating much, if anything.

It is first instructive to see why the Butterfly Process is \emph{not}
\cryptic{1}. 

If we can find a family $\{ \future_0 \}$ such that
$H[\CausalState_1 | \Future_0 = \future_0] \neq 0$, then the total conditional
entropy will be positive and, thus, the machine will not be \cryptic{1}. To
show that this can happen, consider the future
$\future_0 = (0,1,2,4,4,4,\ldots$). It is clear that the
state following $1$ must be $A$. Thus, in order to generate $0$ or $1$ before
arriving at $A$, the state pair $(\CausalState_0, \CausalState_1)$ can be
either $(B,C)$ or $(D,E)$. This uncertainty in $\CausalState_1$ is enough
to break the criterion. And this occurs for the family
$\{ \future_0 \} = \{ 0, 1, \ldots \}$.

To see that the process is \cryptic{2}, notice that the two paths $(B,C)$
and $(D, E)$ converge on $A$. Therefore, there is no uncertainty in
$\CausalState_2$ given this future. It is reasonably straightforward to see
that indeed \emph{any} $(\MeasSymbol_0, \MeasSymbol_1)$ will lead to a unique
causal state. This is because the Butterfly Process is a very limited version
of an $8$-symbol \order{2} Markov process.

Note that the transition matrix is doubly-stochastic and so the stationary
distribution is uniform. The statistical complexity is rather direct in this
case: $\Cmu = \log_2(5)$. We now can calculate $\PC$ using Cor. \ref{cor:OKCC}:
\begin{align*}
\PC & = \PC(2)\\
    & = H[\CausalState_0 | \MeasSymbol_0^2, \CausalState_2]\\
    & = H[\CausalState_0 | \MeasSymbol_0^2]\\
    & = \Pr(01) H[\CausalState_0 | \MeasSymbol_0^2 = 01] + \Pr(12)
H[\CausalState_0 | \MeasSymbol_0^2 = 12] \\
	& ~~~ + \Pr(13) H[\CausalState_0 | \MeasSymbol_0^2 = 13]\\
    & = 2 \frac{1}{4} \frac{1}{5} 1 + 2 \frac{1}{4} \frac{1}{5} 1 + 2
\frac{1}{4} \frac{1}{5} 1\\
    & = \frac{3}{10} ~\mathrm{bits}.
\end{align*}
From \corref{cor:OKCE}, we get an excess entropy of
\begin{align*}
\EE & = \Cmu - \PC(2) \\
    & = \log2(5) - \frac{3}{10} \\
	& \approx 2.0219~\mathrm{bits}.
\end{align*}

For comparison, if we had assumed the Butterfly Process was \cryptic{1}, then
we would have:
\begin{align*}
\EE & = \Cmu - \PC(1) \\
  & = \Cmu - (H[\CausalState_0, \MeasSymbol_0]
	- H[\CausalState_1, \MeasSymbol_0]) \\
  & \approx \log2(5) - (3.3219 - 2.5062) \\
  & = \log2(5) - 0.8156 \approx 1.5063 ~\mathrm{bits}.
\end{align*}
We can see that this is substantially below the true value: a 25\% error.

\subsection{Restricted Golden Mean: \Cryptic{k}}

\begin{figure}[th]
\centering
\includegraphics{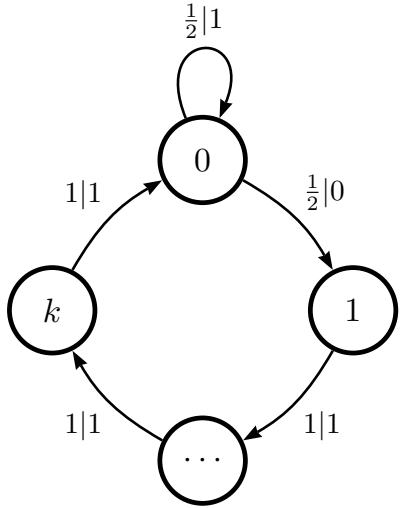}
\caption{\cryptic{k} processes: Restricted Golden Mean Family.}
\label{fig:RestrictedGM}
\end{figure}

Now we turn to illustrate a crypticity-parametrized family of processes,
giving examples of \cryptic{k} processes for any $k$. We call this family the
Restricted Golden Mean as its support is a restriction of the Golden Mean
support. (See \figref{fig:RestrictedGM} for its \eMs.)
The $k=1$ member of the family is exactly the Golden Mean.

It is straightforward to see that this process is \order{k} Markov. Proposition
\ref{prop:MarkovOKC} then implies it is (at most) \cryptic{k}. In order to show
that it is not \cryptic{(k-1)}, consider the case $\future_0 = 1^{k},0,\ldots$.
The first $(k-1)$ $1$s will induce a mixture over states $k$ and $0$. The
following future $\future_k = 1,0,\ldots$ is consistent with both states $k$
and $0$. Therefore, the \crypticity{(k-1)} criterion is not satisfied.
Therefore, it is \cryptic{k}.

For arbitrary $k$, there are $k+1$ causal states and the stationary
distribution is:
\begin{equation*}
\pi = \left( \frac{2}{k+2}, \frac{1}{k+2},
	\frac{1}{k+2}, \ldots, \frac{1}{k+2} \right) ~.
\end{equation*}
The statistical complexity is
\begin{align*}
\Cmu &= \log_2 (k+2) - \frac{2}{k+2} ~.
\end{align*}
For the $k$-th member of the family, we have for the crypticity:
\begin{equation*}
\PC = \PC(k) = \frac{2 k}{k+2} ~.
\end{equation*}
And the excess entropy follows directly from \corref{cor:OKCE}:
\begin{align*}
\EE & = \Cmu - \PC \\
	& =  \log_2 (k+2) - \frac{2(k+1)}{k+2} ~,
\end{align*}
which diverges with $k$. (Calculational details will be provided elsewhere.)

\subsection{Stretched Golden Mean}

The Stretched Golden Mean is a family of processes that does not occupy the
same support as the Golden Mean. Instead of requiring that blocks of 0s are
of length $1$, we require that they are of length $k$. Here, the Markov order
(k) grows, but the cryptic order remains $1$ for all $k$.

Again, it is straightforward to see that this process is \order{k} Markov. To
see that it is \cryptic{1}, first note that if $\MeasSymbol_0 = 1$, then
$\CausalState_1=0$. Next consider the case when $\MeasSymbol_0 = 0$. If the
future $\future_1 = 1, \ldots$, then $\CausalState_1 = k$. Similarly, if the
future $\future_1 = 0^n, 1, \ldots$, then $\CausalState_1 = k-n$. This family
exhibits arbitrary separation between its Markov order and its cryptic
order and so demonstrates that these properties are not redundant.

\begin{figure}[th]
\centering
\includegraphics{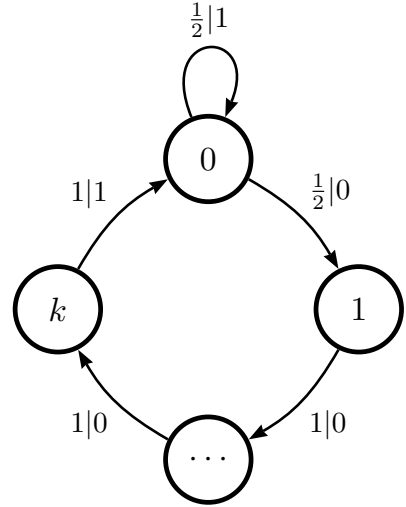}
\caption{\cryptic{k} processes: Stretched Golden Mean Family.}
\label{fig:StretchedGM}
\end{figure}

The stationary distribution is the same as for the Restricted Golden Mean
and so, then, is the statistical complexity. In addition, we have:
\begin{align*}
\PC = \PC(1) &= H[\CausalState_0 | \MeasSymbol_0, \CausalState_1]\\
  &= \hmu ~.
\end{align*}
Consequently,
\begin{align*}
\EE = \Cmu - \PC = \Cmu - \hmu ~.
\end{align*}

\subsection{The Nemo Process: \Cryptic{\infty}}

\begin{figure}[th]
\centering
\includegraphics{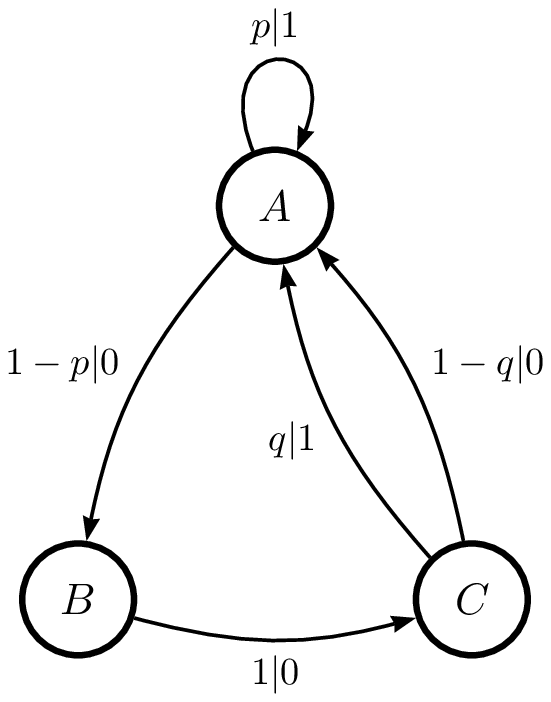}
\caption{The \cryptic{\infty} Nemo Process.
  }
\label{fig:Nemo}
\end{figure}

We close our cryptic process bestiary with a (very) finite-state process that
has infinite crypticity: The three-state Nemo Process. Over no finite-length
sequence will all of the internal state information be present in the
observations. The Nemo Process \eM\ is shown in Fig.~\ref{fig:Nemo}.

Its stationary state distribution is
\begin{align*}
\Pr(\CausalState) \equiv \pi = \frac{1}{3-2p}
\bordermatrix{%
& A & B & C \cr
& 1 & 1-p & 1-p
},
\end{align*}
from which one calculates the statistical complexity:
\begin{align*}
\Cmu &= \log_2(3-2p) - \frac{2(1-p)}{3-2p} \log_2 (1-p) ~.
\end{align*}

The Nemo Process is not a finite-cryptic process. That is, there exists no 
finite $k$ for which $H[\CausalState_k | \Future_0] = 0$. To show this,
we must demonstrate that there exists a family of futures such that for each 
future \mbox{$H[\CausalState_k | \Future_0 = \future] > 0$}.  The family 
of futures we use begins with all $0$s and then has a $1$. Intuitively, 
the $1$ is chosen because it is a synchronizing word for the 
process---after observing a $1$, the \eM\ is 
always in state $A$.  Then, causal shielding will decouple the infinite
future from the first few symbols, thereby allowing us to compute the
conditional entropies for the entire family of futures.

First, recall 
the shorthand:
\begin{align*}
\Pr(\CausalState_k | \Future_0) = 
 \lim_{L\rightarrow\infty} \Pr(\CausalState_k | \MeasSymbol_0^L) ~.
\end{align*}
Without loss of generality, assume $k < L$. Then,
\begin{align*}
\Pr(\CausalState_k | \MeasSymbol_0^L) 
   &= \frac{\Pr(\MeasSymbol_0^k, \CausalState_k, \MeasSymbol_k^L)}
           {\Pr(\MeasSymbol_0^L)} \\
   &= \frac{\Pr(\MeasSymbol_k^L | \MeasSymbol_0^k, \CausalState_k)
\Pr(\MeasSymbol_0^k, \CausalState_k)} 
           {\Pr(\MeasSymbol_0^L)} \\
   &= \frac{\Pr(\MeasSymbol_k^L | \CausalState_k)
\Pr(\MeasSymbol_0^k, \CausalState_k)} 
           {\Pr(\MeasSymbol_0^L)}  ~,
\end{align*}
where the last step is possible since the causal states are
Markovian~\cite{Crut98d},
shielding the past from the future. Each of these quantities is given by:
\begin{align*}
\Pr(\MeasSymbol_k^L = w | \CausalState_k = \causalstate) &=
[T^{(w)}\one]_{\causalstate}\\
\Pr(\MeasSymbol_0^k=w, \CausalState_k = \causalstate) &= [\pi
T^{(w)}]{}_\causalstate \\
\Pr(\MeasSymbol_0^L=w) &= \pi T^{(w)} \one ~.
\end{align*}
where $T^{(w)} \equiv T^{(\meassymbol_0)}T^{(\meassymbol_1)}\cdots
T^{(\meassymbol_{L-1})}$, $\one$ is a column vector of $1$s, and
$T^{(\meassymbol)}_{\causalstate\causalstateprime} = 
\Pr(\CausalStatePrime = \causalstateprime, 
    \MeasSymbol=\meassymbol | \CausalState=\causalstate)$.
To establish $H[\CausalState_k | \Future_0] > 0$ for any $k$, we rely on using
values of $k$ that are multiples of three. So, we concentrate on the following
for $n = 0,1,2,\ldots$:
\begin{align*}
H[\CausalState_{3n} | \MeasSymbol_0^{3n+1} = 0^{3n}1, \Future_{3n+1}] > 0 ~.
\end{align*}
Since $1$ is a synchronizing word, we can greatly simplify the conditional
probability distribution.  First, we freely include the synchronized causal 
state $A$ and rewrite the conditional distribution as fraction:
\begin{align*}
& \Pr(\CausalState_{3n} | \MeasSymbol_0^{3n+1} = 0^{3n}1, \Future_{3n+1})\\
&= \Pr(\CausalState_{3n} | \MeasSymbol_0^{3n+1} = 0^{3n}1, \CausalState_{3n+1}=A, \Future_{3n+1})\\
&= \frac{\Pr(\CausalState_{3n}, \MeasSymbol_0^{3n+1} = 0^{3n}1, \CausalState_{3n+1}=A, \Future_{3n+1})}
        {\Pr(\MeasSymbol_0^{3n+1} = 0^{3n}1, \CausalState_{3n+1}=A, \Future_{3n+1})} ~.
\end{align*}
Then, we factor everything except $\Future_{3n+1}$ out of the numerator 
and make use of causal shielding to simplify the conditional. For example,
the numerator becomes:
\begin{align*}
& \Pr(\CausalState_{3n}, \MeasSymbol_0^{3n+1} = 0^{3n}1, \CausalState_{3n+1}=A, \Future_{3n+1})\\
&= \Pr(\Future_{3n+1}|\CausalState_{3n}, \MeasSymbol_0^{3n+1} = 0^{3n}1, \CausalState_{3n+1}=A) \\
&\qquad \times \Pr(\CausalState_{3n}, \MeasSymbol_0^{3n+1} = 0^{3n}1, \CausalState_{3n+1}=A)\\
&= \Pr(\Future_{3n+1}|\CausalState_{3n+1}=A) \\
&\qquad \times 
\Pr(\CausalState_{3n}, \MeasSymbol_0^{3n+1} = 0^{3n}1, \CausalState_{3n+1}=A)\\
&= \Pr(\Future_{3n+1}|\CausalState_{3n+1}=A) \Pr(\CausalState_{3n}, \MeasSymbol_0^{3n+1} = 0^{3n}1)~ .
\end{align*}
Similarly, the denominator becomes:
\begin{align*}
& \Pr(\MeasSymbol_0^{3n+1} = 0^{3n}1, \CausalState_{3n+1}=A, \Future_{3n+1})\\
&= \Pr(\Future_{3n+1}|\CausalState_{3n+1}=A) \Pr(\MeasSymbol_0^{3n+1} = 0^{3n}1)~ .
\end{align*}
Combining these results, we obtain a finite form for the entropy of $S_{3n}$ 
conditioned on a family of infinite futures, first noting:
\begin{align*}
\Pr(\CausalState_{3n} | \MeasSymbol_0^{3n+1} = 0^{3n}1, \Future_{3n+1}) 
    = \Pr(\CausalState_{3n} | \MeasSymbol_0^{3n+1} = 0^{3n}1) ~.
\end{align*}
Thus, for all $\future_{3n+1}$, we have:
\begin{align*}
& H[\CausalState_{3n} | \MeasSymbol_0^{3n+1} = 0^{3n}1, 
                        \Future_{3n+1}=\future_{3n+1}] \\
&= H[\CausalState_{3n} | \MeasSymbol_0^{3n+1} = 0^{3n}1] ~.
\end{align*}

Now, we are ready to compute the conditional entropy for the entire family.
First, note that $T^{(0)}$ raised to the third power is a diagonal matrix with 
each element equal to $(1-p)(1-q)$. Thus, for $j=1,2,3\ldots$:
\begin{align*}
\bigl[T^{(0)}\bigr]^{3j}_{\causalstate\causalstate} &= (1-p)^j(1-q)^j ~.
\end{align*}
Using all of the above relations, we can easily calculate:
\begin{align*}
\Pr(\CausalState_{3n} | \MeasSymbol_0^{3n+1} = 0^{3n+1}1)
&= \frac{1}{3-2p}
\bordermatrix{%
& A & B & C \cr
& p & 0 & q(1-p)
} ~.
\end{align*}
Thus, for $p,q \in (0,1)$, we have:
\begin{align*}
&H[\CausalState_{3n} | \Future_0] \\
    &\geq H[\CausalState_{3n} | \MeasSymbol_0^{3n+1} = 0^{3n}1, 
                                \Future_{3n+1}] \\
    &= \sum_{\future_{3n+1}} \Pr\left(
                \MeasSymbol_0^{3n+1} = 0^{3n}1, 
                \Future_{3n+1}=\future_{3n+1}
               \right) \\
    &\quad\times H[\CausalState_{3n} | \MeasSymbol_0^{3n+1} = 0^{3n}1, \Future_{3n+1} = \future_{3n+1}] \\
    &= H[\CausalState_{3n} | \MeasSymbol_0^{3n+1} = 0^{3n}1] \\
    &\quad\times 
       \sum_{\future_{3n+1}} \Pr\left(
                \MeasSymbol_0^{3n+1} = 0^{3n}1, 
                \Future_{3n+1}=\future_{3n+1}
             \right) \\
    &= H[\CausalState_{3n} | \MeasSymbol_0^{3n+1}=0^{3n}1] \Pr(\MeasSymbol_0^{3n+1}=0^{3n}1)\\
    &= \left(\frac{p}{3-2} \log_2 \frac{3-2p}{p} + 
       \frac{q(1-p)}{3-2p} \log_2 \frac{q(1-p)}{3-2p}\right)\\
    &\quad \times [(1-p)(1-q)]^{3n} \\
    &> 0 ~.
\end{align*}
So, any time $k$ is a multiple of three, $H[S_k|\Future_0] > 0$.
Finally, suppose $(k \mod 3) = i$, where \mbox{$i\neq 0$}. That is, suppose $k$ 
is not a multiple of three. By \lemref{lem:hSkgF}, 
$H[\CausalState_k | \Future_0] \geq H[\CausalState_{k+i} | \Future_0]$ and,
since we just showed that the latter quantity is always strictly greater 
than zero, we conclude that $H[\CausalState_k|\Future_0] > 0$ for every 
value of $k$.

The above establishes that the Nemo Process does not satisfy the \crypticity{k}
criterion for any finite $k$. Thus, the Nemo process is \cryptic{\infty}. This
means that we cannot make use of the \cryptic{k} approximation to calculate
$\PC$ or $\EE$.

Fortunately, the techniques introduced in \refcite{Crut08a}
and \refcite{Crut08b} do not rely on an approximation method. To avoid
ambiguity denote the statistical complexity we just computed as $\FutureCmu$.
When the techniques are applied to the Nemo Process, we find that the process
is causally reversible ($\FutureCmu = \PastCmu$) and has the following
forward-reverse causal-state conditional distribution:
\begin{align*}
\Pr(\FutureCausalState|\PastCausalState) &= \frac{1}{p + q - pq}
\bordermatrix{
  & A & B & C \cr
D & p & 0 & q(1-p) \cr
E & 0 & q & p(1-q) \cr
F & q & p(1-q) & 0 \cr
}
 ~.
\end{align*}
With this, one can calculate $\EE$, in closed-form, via:
\begin{align*}
\EE = \FutureCmu - H[\FutureCausalState|\PastCausalState] ~.
\end{align*}
(Again, calculational details will be provided elsewhere.)

\section{Conclusion}

Calculating the excess entropy $I[\Past ; \Future]$ is, at first blush, a
daunting task. We are asking for a mutual information between two infinite sets
of random variables. Appealing to $\EE = I[\CausalState ; \Future]$, we use the
compact representation of the \eM\ to reduce one infinite set (the past) to a
(usually) finite set. A process's \crypticity{k} captures something similar
about the infinite set of future variables and allows us to further compact
our form for excess entropy, reducing an infinite variable set to a
finite one. The resulting stratification of process space is a novel way of
thinking about its \emph{structure} and, as long as we know which stratum we
lie in, we can rapidly calculate many quantities of interest.

Unfortunately, in the general case, one will not know a priori a process's
crypticity order. Worse, as far as we are aware, there is no known finite
method for calculating the crypticity order. This strikes us as an interesting
open problem and challenge.

If, by construction or by some other means, one does know it, then, as we
showed, crypticity and $\EE$ can be calculated using the crypticity expansion.
Failing this, though, one might consider using the expansion to search for the
order. There is no known stopping criterion, so this search may not find $k$ 
in finite time. Moreover, the expansion is a calculation that grows
exponentially in computational complexity with crypticity order, as we noted.
Devising a stopping criterion would be very useful to such a search.

Even without knowing the \crypticity{k}, the expansion is often still useful. 
For use in estimating $\EE$, it provides us with a bound from above. This is 
complementary to the bound below one finds using the typical expansion 
$\EE(L) = H[\MeasSymbol_0^L] - \hmu L$ \cite{Crut01a}. Using these upper and
lower bounds, one may determine that for a given purpose, the estimate of
$\PC$ or $\EE$ is within an acceptable tolerance.

The crypticity hierarchy is a revealing way to carve the space of processes
in that it concerns how they hide internal state information from an observer. 
The examples were chosen to illustrate several features of this new view. The 
Even Process, a canonical example of \order{\infty} Markov, resides instead at
the very bottom of this ladder. The two example families show us how \cryptic{k}
is neither a parallel nor independent concept to \order{R} Markov. Finally, we
see in the last example an apparently simple process with \crypticity{\infty}. 

The general lesson is that internal state information need not be immediately
available in measurement values, but instead may be spread over long measurement
sequences. If a process is \cryptic{k} and $k$ is finite, then internal state
information is accessible over sequences of length $k$. The
existence, as we demonstrated, of processes that are \cryptic{\infty} is rather
sobering. (The Appendix comments on what happens when one fails to appreciate
this.) Interpreted as a statement of the impossibility of extracting state
information, it reminds us of earlier work on hidden spatial dynamical systems
that exhibit a similar encrypting of internal structure in observed spacetime
patterns \cite{Crut91e}.

Due to the exponentially growing computational effort to search for the
crypticity order and, concretely, the existence of \cryptic{\infty} processes,
the general theory introduced in \refcite{Crut08a} and \refcite{Crut08b} is
seen to be necessary. It allows one to directly calculate $\EE$ and crypticity
and to do so efficiently.

\section*{Acknowledgments}

Chris Ellison was partially supported on a GAANN fellowship. The Network
Dynamics Program funded by Intel Corporation also partially supported this
work.

\appendix

\section{Appendix: Crypticity Untamed}

Recently, Ref. \cite{Wies09a} asserted that a process's $\EE$ can be obtained
from its \eM\ using the following expression:
\begin{equation*}
\EE = \Cmu - I_{erased} ~,
\end{equation*}
where $I_{erased} = H[\CausalState_0,\MeasSymbol_0]
- H[\CausalState_1,\MeasSymbol_0]$.
Though renamed, $I_{erased}$ is the crypticity of \refcite{Crut08a}.
However, as we showed in the main development, it is $\FuturePC(1)$ and so the
above expression is valid only for \cryptic{0} and \cryptic{1} processes.

Ref. \cite{Wies09a} considered only the Even and Golden Mean
Processes. These, as we saw, are \cryptic{0} and \cryptic{1} and so it is no
surprise that the expression worked. Indeed, their low-order crypticity is why
closed-form expressions for their excess entropies have been known for quite
some time, prior to the recent developments.

In short, the claims in Ref. \cite{Wies09a} are incorrect. The implication there
that all \eMs\ are \cryptic{1} is also. The examples we gave show how wrong
such an approximation can be. We showed how large the errors can grow. The
full theory of \refcite{Crut08a} and \refcite{Crut08b} is required. The
richness of the space of processes leads us to conjecture that it will suffer
no shortcuts.
\bibliography{ref,chaos}

\end{document}